\documentclass[iop]{emulateapj}

\usepackage{natbib}
\begin{document}

\title{Investigation of the Progenitors of the Type Ia Supernovae Associated With the LMC Supernova Remnants 0505-67.9 and 0509-68.7}
\author{Ashley Pagnotta}
\affil{Department of Astrophysics, American Museum of Natural History, New York, NY 10024}
\author{Bradley E. Schaefer}
\affil{Department of Physics and Astronomy, Louisiana State University, Baton Rouge, LA 70803}

\begin{abstract}
Although Type Ia supernovae have been heavily scrutinized due to their use in making cosmological distance estimates, we are still unable to definitively identify the progenitors for the entire population. While answers have been presented for certain specific systems, a complete solution remains elusive. We present observations of two supernova remnants (SNRs) in the Large Magellanic Cloud, SNR 0505-67.9 and SNR 0509-68.7, for which we have identified the center of the remnant and the 99.73\% containment central region in which any companion star left over after the supernova must be located. Both remnants have a number of potential ex-companion stars near their centers; all possible single and double degenerate progenitor models remain viable for these two supernovae. Future observations may be able to identify the true ex-companions for both remnants.
\end{abstract}

\keywords{supernova remnants}

\section{Type Ia Supernova Progenitor Searches}
\label{sec:intro}
The Type Ia supernova (SN Ia) progenitor problem has existed for decades but has recently received more attention due to the use of SN Ia-derived distances in cosmology to discover the acceleration of the expansion of the universe (\citealp{riess1998a, perlmutter1999a}; see \mbox{\citet{wang2012a}} and \mbox{\citet{maoz2014a}} for recent reviews of the SN Ia progenitor problem). It is widely accepted that SNe Ia result from the explosion of a carbon-oxygen white dwarf (WD) that has reached the Chandrasekhar mass limit due to interaction with a companion star, either through accretion or a merger \mbox{{\citep{nomoto1997a}}}, but the identity of that companion has long been a mystery. Many companion/progenitor possibilities have been proposed, and they can be divided into two main classes, the double degenerates (DDs), which consist of two carbon-oxygen WDs that inspiral and combine to explode, and the single degenerates (SDs), which consist of a non-degenerate companion star that donates mass to the WD. The current possible SD progenitor candidates are recurrent novae \mbox{(RNe; \citealp{hachisu2001a})}, symbiotic stars \mbox{\citep{hachisu1999a}}, supsersoft X-ray sources (SSSs; \citealp{hachisu1999b,langer2000a,han2004a}), and helium donor stars \mbox{\citep{wang2009a}}. Additionally, spin-up/spin-down models have been proposed by \mbox{\citet{justham2011a} and \citet{di-stefano2011a}} as modifications of the standard possible SD progenitors to explain the lack of hydrogen in the spectra of most SNe Ia as well as the lack of any evidence of interaction with an SD companion, i.e., lack of detection in either radio or X-rays.

One way to identify the companion stars is to look near the center of local Type Ia supernova remnants (SNRs) after the explosion for evidence of a leftover ex-companion \citep{ruiz-lapuente1997a,canal2001a}. All of the SD models will leave behind the non-degenerate star, which will remain relatively close to the center of the remnant and be detectable for centuries after the explosion. If there are no companion stars present to sufficiently deep limits, then all of the SD models can be ruled out and the DD model becomes the only remaining possibility.

Each of the potential SN Ia progenitor systems listed above leaves behind a telltale post-supernova signature. In the DD scenario, no ex-companion will be left over, as both WDs are destroyed in the explosion. In the SD scenario, each possible system leaves behind a certain type of ex-companion. These possible ex-companion stars are listed in Table 1 of \mbox{\citet{schaefer2012a}}, which also gives a summary of their intrinsic properties as well as their expected apparent magnitude given their location at the distance of the Large Magellanic Cloud (LMC). 

Each of the possible SD ex-companions has a characteristic velocity after the SN explosion. This velocity is due to two main factors: the orbital velocity prior to the explosion and the kick from the supernova itself. The orbital velocities are calculated for each type of ex-companion (red giant, subgiant, $M>1.16 M_\odot$ main-sequence star, and Helium star) and assume that the companion was filling its Roche Lobe just prior to the SN. We also account for the orbital velocity of the exploding WD and correspondingly the remnant. The kicks from SNe Ia are relatively small \citep{canal2001a,marietta2000a,pan2010a}. From \citet{canal2001a},  \citet{marietta2000a}, and \mbox{\cite{wang2009b}}, the average post-explosion velocities are 100 km s{$^{-1}$}, 250 km s{$^{-1}$}, 390 km s{$^{-1}$}, and 400 km s{$^{-1}$} for red giants, subgiants, 1.16 $M_\odot$ main-sequence ex-companions, and Helium stars, respectively.

The proximity of the actual SN explosion will affect the ex-companion beyond just imparting a kick. The loosely bound outer layers of evolved companions will likely be stripped off by the blast and expanding shell of the SN, but main-sequence stars and other possible ex-companions with high surface gravity, such as Helium star or already-stripped red giant cores, will not undergo any significant stripping. Detailed simulations indicate that the effect will not greatly change the location of the high surface gravity stars on the H-R diagram \citep{marietta2000a,pan2010a,podsiadlowski2003a,pan2014a}. In the subgiant case, calculations by \citet{marietta2000a} show that the ex-companion will likely be two orders of magnitude brighter, although in rare cases the remaining subgiant could be up to ten times less luminous if there is low energy deposition. Newer simulations by \citet{shappee2013a} that consider the effect of ablation show that the energy deposited onto the companion will cause it to be significantly overluminous, up to $50-60L_\odot$ 100 years after the explosion and $15-20L_\odot$ 1000 years later. Regardless of which set of simulations proves to be more correct, the takeaway result is that being next to an exploding WD will almost certainly not dim the ex-companion\textemdash it will survive the explosion looking either quite similar to how it did before or much brighter.

In addition to a possible increase in luminosity, true ex-companions may have additional identifying features. All of the SD models consist of a tidally locked, relatively tight binary just before the explosion. Post-explosion, the ex-companion will still be moving quickly and rotating rapidly, two effects that can be measured using high-resolution spectroscopy. The high velocity can be sought by measuring the radial velocity or, for Galactic remnants, proper motion. Depending on the direction of the ex-companion's motion, however, it is possible that we will be unable to easily detect this high velocity from Earth. High radial velocity can be detected in spectroscopic observations, but there are indications that it is possible for the radial velocity to drop sharply during the explosion \citep{pan2012b}. Both of these effects, if observed, can identify an ex-companion; their absence for any given star, however, does not preclude it from being the ex-companion. Another observational feature to look for is the presence of blue-shifted iron lines in the spectrum of the possible ex-companion, which would indicate that the star is in fact located within the expanding SNR, and not in front of or behind it \citep{ozaki2006a}. Again, however, the absence of these blueshifted lines is not proof that a given star is not the ex-companion, because their presence depends on the explosion model and the ionization state of the ejecta \citep{ozaki2006a}.

\citet{ruiz-lapuente2004a} used this method of searching for ex-companion stars within historical SNRs for the first time with the Galactic remnant of SN 1572 (Tycho's SN). They identified a G2 IV subgiant star, which they named Star G, as the ex-companion, implying an RN or SSS progenitor. Star G was identified based on its proximity to the center of the SNR as well as its high proper motion, but these properties are still disputed \citep{fuhrmann2005a,ihara2007a,kerzendorf2009a,gonzalez-hernandez2009a,kerzendorf2013a}. The spectral analysis performed by \citet{gonzalez-hernandez2009a} shows that nickel and cobalt are anomalously overabundant in the atmosphere of Star G, which suggests it was blasted by the exploding WD during the SN.  \citet{gonzalez-hernandez2012a} and \citet{kerzendorf2012a} have also examined the remnant of SN 1006, the only other confirmed Ia SNR in our Galaxy. Both papers rule out red giant and subgiant ex-companion stars, because none of the evolved stars near the center of the SNR (as seen from Earth) at a radial distance even somewhat coincident with the remnant show any indication of being an ex-companion, and \citet{kerzendorf2012a} state that they find no stars ``consistent with the traditional accretion scenario", i.e., all SD progenitor models. There are two other possible Ia SNRs in the Galaxy, the remnant of SN 1604 (Kepler) and RCW 86. Although the type of SN 1604 has not been confirmed by light echoes, there are indications that it was a Type Ia, with the latest evidence coming from Iron K-shell emission from the remnant \mbox{\citep{yamaguchi2014a}}, and investigations of the central stars are ongoing (\mbox{\citealp{kerzendorf2014a}} and R. Sankrit 2014, private communication). RCW 86 may \mbox{\citep{williams2011b}} or may not \mbox{\citep{chin1994a,schaefer1995a}} be from a supernova associated with the Chinese guest star of AD 185. If it is, its large size, age, and uncertain origin combine to produce a situation in which it is unlikely any useful progenitor information can be gleaned (see Section {\ref{sec:0505}} for further discussion on the impact of age especially).

Poorly known distances to both the remnants and the nearby stars make Galactic searches very difficult, as it is often not clear if stars which appear to be near the remnant center are actually located the same distance from Earth as the remnant itself. To combat this problem, we extended this method to the LMC, which has a well known distance \citep{freedman2001a,schaefer2008a}, thorough extinction maps \citep{zaritsky2004a} and, in general, less crowded star fields than the Milky Way. Although LMC SNRs are significantly farther away, all possible companion stars are still observable, ranging from $16 \le V \le 22.7$; see Table 1 of \mbox{\citet{schaefer2012a}} for details.

Our first LMC target was SNR 0509-67.5, which is the remnant of a 1991T-type SN Ia that occurred $400 \pm 50$ years ago \citep{badenes2009a,hughes1995a,rest2005a,rest2008a}. The Hubble Space Telescope ({\it HST}) archive contained WFPC2 H$\alpha$ images (PI J. P. Hughes, Rutgers) and WFC3 {\it BVI} images (PI K. S. Noll, Hubble Heritage Program) that provided excellent coverage of the area. We used three independent methods to obtain the geometric center of the remnant and then calculated a 1.4" offset between the geometric center and the actual explosion site due to an enhancement of circumstellar dust in the southwest quadrant of the remnant. We then searched for possible ex-companion stars in a circular region centered on the explosion site with a 1.43" radius corresponding to the extreme 3$\sigma$ ($99.73 \%$ containment) distance that any possible ex-companion could have traveled. This central region was devoid of point sources to the limiting magnitude of $V=26.9$, which corresponds to $M_\mathrm{V}=+8.4$ at the distance of the LMC. The lack of any ex-companions visible in this region rules out all SD progenitor models, leading us to conclude that the only possible progenitor for this supernova is a DD system \citep{schaefer2012a}. This was the first unambiguous result for any known SN Ia progenitor. Although there are no point sources in the central region of the SNR, there is some nebulosity visible, which a GMOS long-slit spectrum showed is a background galaxy, unrelated to the supernova remnant \citep{pagnotta2014a}.

Our second LMC target was SNR 0519-69.0, which was produced by a normal SN Ia that exploded $600 \pm 200$ years ago (\citealp{rest2005a}; A. Rest 2010, private communication). There were archival {\it HST} ACS images of this remnant in both the H$\alpha$ and $V$-band filters, originally taken in 2011 (PI J. P. Hughes, Rutgers). The outer edge of the shell shows a number of small-scale variations, but overall it is nearly symmetric as long as the faint outer arc in the northeast quadrant of the remnant is considered. Because of this, we considered the explosion site to be at the geometric center. Again we constructed a central error region around the explosion site that corresponds to 3$\sigma$ containment of all possible ex-companion stars. Compared to SNR 0509-67.5, the central error region for SNR 0519-69.0 is significantly larger due to the asymmetries in the shell as well as the longer time since explosion. The 4.7" central region for SNR 0519-69.0 contains 127 stars, including 27 main-sequence stars that are bright enough that they could be ex-companions from a SSS binary. There are no post-main-sequence stars in the central error region; the nearest red giant and subgiant are 6.0" and 7.4" from the center, respectively. We therefore concluded that the SN Ia that created SNR 0519-69.0 could have only come from an SSS or DD progenitor system \cite{edwards2012a}.

Here we describe similar searches near the centers of the two remaining confirmed SNe Ia in the LMC, SNR 0505-67.9 and SNR 0509-68.7. We find that both remnants have potential evolved ex-companion stars and thus conclude that all possible SN Ia progenitor models are still possibilities for each of these two SNe.

\section{LMC SNR 0505-67.9}
\label{sec:0505}
\begin{figure*}
\centering
\epsscale{1.0}
\plotone{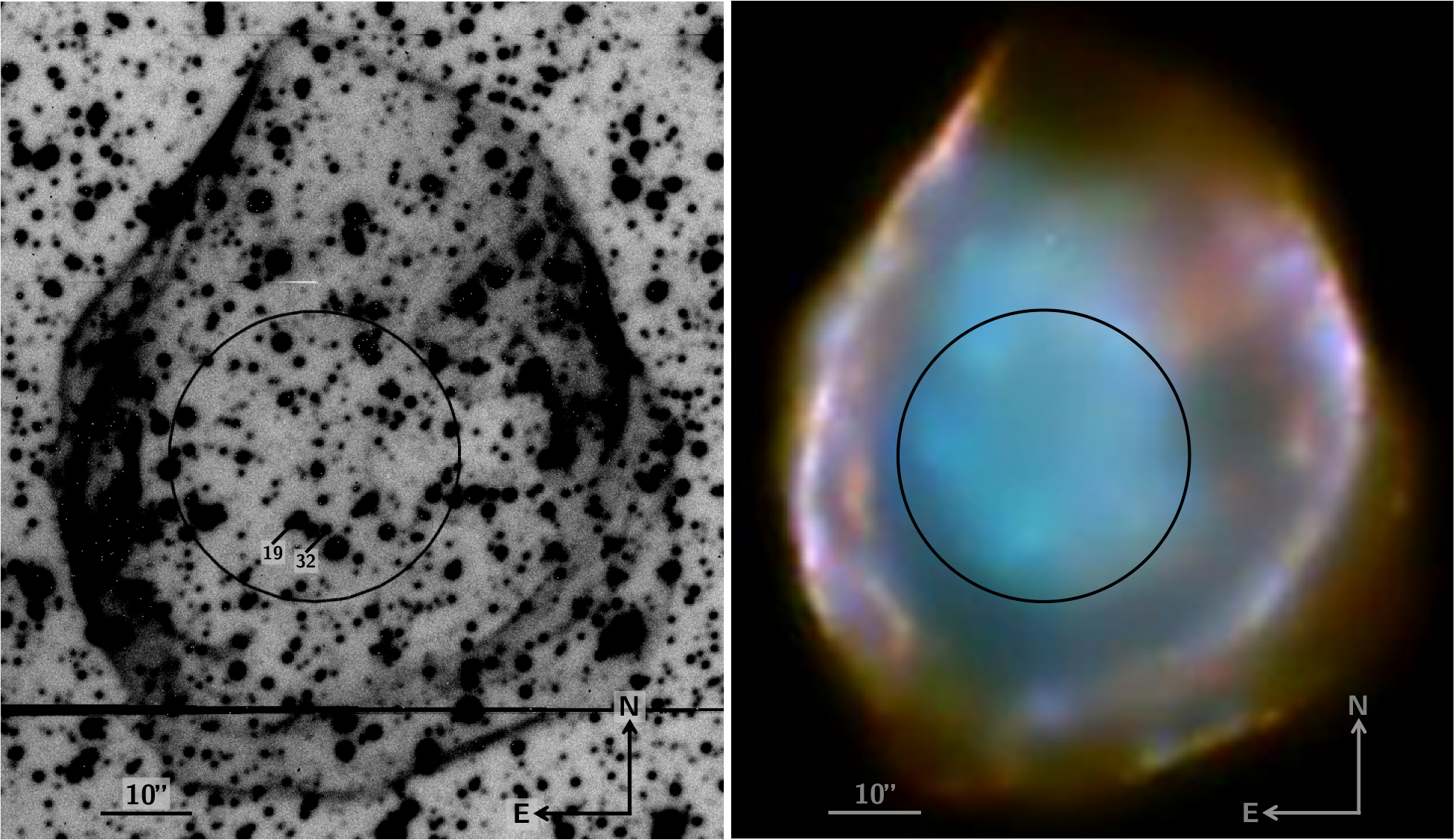}
\caption{Identically cropped images of LMC SNR 0505-67.9 (DEM L71); on the left is the combined {\it g'r'i'}+H$\alpha$ Gemini image, and on the right is the Chandra color image, which is  a combination of 0.3-0.7 keV (red), 0.7-1.1 keV (green), and 1.1-2.4 keV (blue) observations. The Gemini images were taken with the GMOS imager on the 8.1m Gemini South Telescope. (The dark horizontal line running across the bottom of the Gemini image is an artifact due to a nearby bright star.) The Chandra image on the right shows the complicated X-ray structure of the remnant. To measure the remnant center, we used four different gas regions visible in X-ray: the extreme outer edge, the rim of the outer shell, the edge of the inner region, and the central minimum. This allows us to get a more accurate measurement of the center as well as a better understanding of the errors on that measurement. The 15.8$\arcsec$ 99.73\% containment central region is marked on both images. The error circle is very large, because the remnant is both old (4360 years) and asymmetric, and therefore it contains a number of possible ex-companion stars. Stars 19 and 32 are marked; star 19 is the nearest red giant to the center, and star 32 is the nearest subgiant. (The numbers correspond to their identifications in Table \ref{tab:0505mags}.) Future spectroscopy may allow us to identify which of the central stars are likely to be the actual ex-companion, if any.   (X-ray image source: The Chandra Supernova Remnant Catalog (http://hea-www.harvard.edu/ChandraSNR/), used with permission)}
\label{fig:0505diptych}
\end{figure*}

The third SN Ia remnant in the LMC we consider is SNR 0505-67.9 (DEM L71). There are no light echoes for this 4360-year old remnant, but X-ray spectra show that it is Balmer-dominated and has enhanced Fe abundances, both of which indicate that it is from an SN Ia \citep{ghavamian2003a,hughes1998a}. There are no {\it HST} observations of this remnant, so we obtained Gemini South GMOS images \citep{hook2004a} on 2011 September 21. The combined {\it g'r'i'}+H$\alpha$ image, with the central error region marked, can be seen in the left panel of Figure \ref{fig:0505diptych}.

We used our GMOS H$\alpha$ images in combination with Chandra X-ray images (\citealp{hughes2003a,rakowski2003a}, obtained via the Chandra SNR Catalog\footnote{http://hea-www.harvard.edu/ChandraSNR/}) and the perpendicular bisector method described in Supplementary Information Section 2 of \citet{schaefer2012a} to identify the geometric center of the remnant. For the X-ray images, we used four different gas regions to locate the center: the extreme outer edge, the rim of the outer shell, the edge of the inner region, and the central minimum. These regions can be seen in the Chandra X-ray color image shown in the right panel of Figure \ref{fig:0505diptych}. All of the center measurements can be seen in Table \ref{tab:0505centers}. The final explosion site is 05:05:42.71, -67:52:43.5 (J2000).

There is an apparent asymmetry in this remnant, in the North direction, where it appears as though the expanding remnant has ``punched through" an area of low interstellar mass (ISM) density, or perhaps has been slowed down in the East and West directions. Spitzer 24{$\mu$} imaging \mbox{\cite{seok2013a}} shows emission in roughly the East and West directions of the remnant, indicating enhanced ISM or pre-existing dust. Since the dust distribution is relatively symmetric across both the North-South and East-West axes, we do not apply any overall offset between the position of the geometric center and the explosion site, but we do account for the asymmetry in general by including a higher measurement uncertainty on the position of the explosion, which increases the size of the central region.

\begin{figure}
\centering
\epsscale{1.0}
\plotone{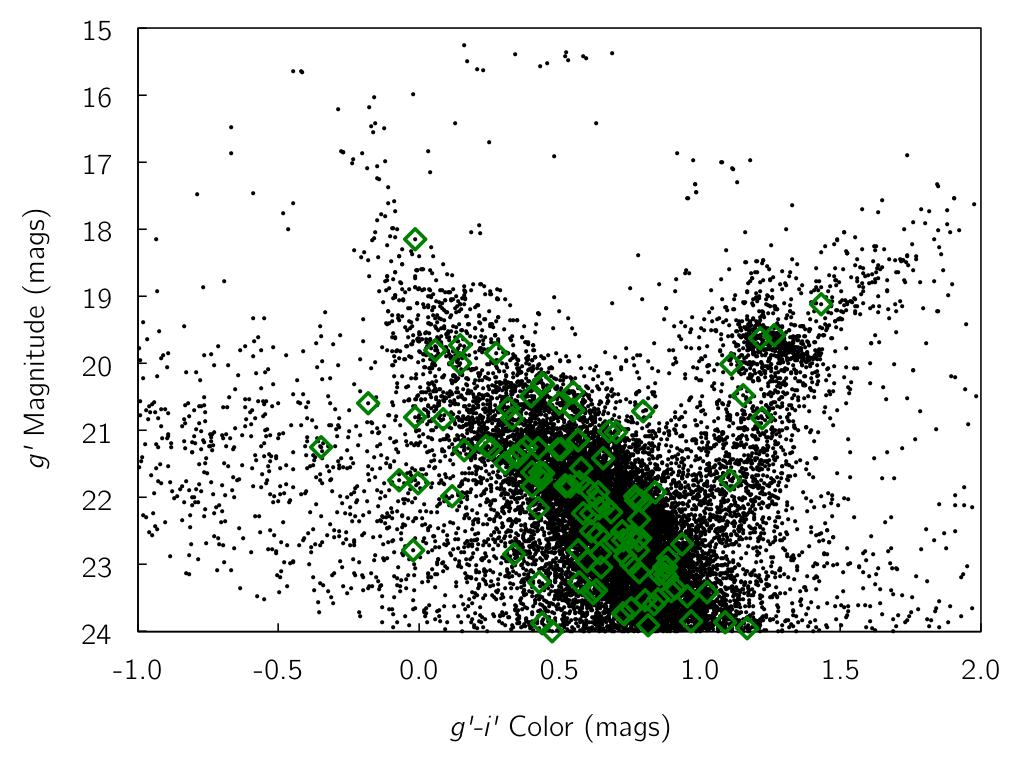}
\caption{$g'-i'$ Color-Magnitude Diagram for SNR 0505-67.9. This $g'-i'$ color-magnitude diagram constructed from our Gemini GMOS observations can be used to classify stars in the field of SNR 0505-67.9. No error bars are plotted on this figure, but a full listing of the photometry can be found in Table \ref{tab:0505mags}; the average $1\sigma$ errors on $g'$ and $g'-i'$  are $\pm 0.08$ and $\pm 0.12$, respectively. The stars located within the 99.73\% containment central region of the supernova remnant are highlighted with green diamonds. There are possible ex-companion stars of all types (main-sequence, subgiant, and red giant) located within the central region.}
\label{fig:0505cmd}
\end{figure}

SNR 0505-67.9 is the oldest of the LMC Ia SNRs by far, at $4360 \pm 290$ years \citep{ghavamian2003a}. During this time, any ex-companion star could have moved a great distance, so the final 99.73\% containment circle is very large, at $15.8 \arcsec$. Because of this, there are quite a lot of stars located within the central region, including six red giants, two possible subgiants, and a number of main-sequence stars bright enough to have come from SSSs. The color-magnitude diagram for all of the stars in the SNR 0505-67.9 field can be seen in Figure \ref{fig:0505cmd}, for which the magnitudes were calibrated using \citet{zaritsky2004a}; the stars located within the central error circle are highlighted with green diamonds and listed in Table \ref{tab:0505mags}. For this remnant, we cannot exclude any progenitor type; all SD and DD progenitors are still possibilities. Further observations of the potential ex-companion stars within the central region, namely high resolution spectra, are needed to identify any likely ex-companions. It may be difficult, however, to ever obtain any conclusive answers for this remnant, due to its age. To do complete follow-up on the large number of stars within the central region would take correspondingly large amounts of telescope and analysis time, and at least one of the identifying characteristics of an ex-companion, the presence of enhanced heavy elements, is likely no longer detectable due to normal mixing processes in the outer atmosphere of any potential ex-companion. In general, the method of searching for ex-companion stars in historical SNRs is best applied to younger remnants; it is likely that all older SNRs will have these same difficulties. It is not, however, impossible to obtain an answer, so investigation and follow-up should be continue to be attempted for all known nearby Ia SNRs.

\begin{centering}
\begin{deluxetable*}{lllll}
\tabletypesize{\scriptsize}
\tablewidth{0pc}
\tablecaption{Positions in SNR 0505-67.9}
\tablehead{
	\colhead{Position} & 
	\colhead{RA (J2000)} & 
	\colhead{Dec. (J2000)} & 
	\colhead{Radius ($\arcsec$)} & 
	\colhead{Confidence}
	}
\startdata
Geometric center in H$\alpha$	&	05:05:41.77	&	-67:52:42.5	&	0.7	&	1$\sigma$	\\
Geometric center in X-ray (0.7-1.1 keV), extreme outer edge	&	 05:05:41.89	&	 -67:52:42.1	&	2.0	&	1$\sigma$	\\
Geometric center in X-ray (0.7-1.1 keV), rim of outer shell	&	 05:05:42.27	&	 -67:52:40.3	&	2.0	&	1$\sigma$	\\
Geometric center in X-ray (0.7-1.1 keV), edge of inner region	&	 05:05:42.46	&	 -67:52:37.7	&	1.0	&	1$\sigma$	\\
Geometric center in X-ray (0.7-1.1 keV), central minimum	&	 05:05:43.00	&	 -67:52:38.9	&	2.0	&	1$\sigma$	\\
Combined geometric center of SNR	&	05:05:42.71	&	-67:52:43.5	&	3.2	&	1$\sigma$	\\
Site of explosion	&	05:05:42.71	&	-67:52:43.5	&	3.2	&	1$\sigma$	\\
Main-sequence ex-companion	&	05:05:42.71	&	-67:52:43.5	&	15.8	&	3$\sigma$	\\
\enddata
\label{tab:0505centers}
\end{deluxetable*}
\end{centering}

\begin{centering}
\begin{deluxetable*}{lllllrl}
\tabletypesize{\scriptsize}
\tablewidth{0pc}
\tablecaption{Stars Inside Central Region for SNR 0505-67.9}
\tablehead{
	\colhead{Star} 				& 
	\colhead{RA (J2000)} 		& 
	\colhead{Dec. (J2000)} 		& 
	\colhead{$\Theta$ ($\arcsec$)}	& 
	\colhead{$g'$ (mag)} 			& 
	\colhead{$g'-i'$ (mag)} 	& 
	\colhead{Comments}
	}
\startdata
1	&	05:05:42.656	&	-67:52:41.97	&	1.516504735	&	$22.23 \pm 0.07$	&	$0.61 \pm 0.11$	&	...	\\
2	&	05:05:42.700	&	-67:52:45.18	&	1.719439881	&	$20.44 \pm 0.07$	&	$0.55 \pm 0.10$	&	Brightest star in central region	\\
3	&	05:05:42.941	&	-67:52:45.35	&	2.319217252	&	$22.64 \pm 0.08$	&	$0.78 \pm 0.11$	&	...	\\
4	&	05:05:42.185	&	-67:52:42.17	&	3.192107543	&	$23.04 \pm 0.08$	&	$0.87 \pm 0.11$	&	...	\\
5	&	05:05:42.524	&	-67:52:40.10	&	3.507364291	&	$22.90 \pm 0.08$	&	$0.89 \pm 0.11$	&	...	\\
6	&	05:05:42.019	&	-67:52:44.64	&	4.030612992	&	$23.39 \pm 0.08$	&	$0.90 \pm 0.11$	&	...	\\
7	&	05:05:42.586	&	-67:52:48.29	&	4.868467449	&	$21.56 \pm 0.07$	&	$0.58 \pm 0.10$	&	...	\\
8	&	05:05:42.726	&	-67:52:38.50	&	4.961735208	&	$23.88 \pm 0.11$	&	$0.44 \pm 0.17$	&	...	\\
9	&	05:05:43.496	&	-67:52:46.10	&	5.196734656	&	$22.60 \pm 0.08$	&	$0.79 \pm 0.11$	&	...	\\
10	&	05:05:41.984	&	-67:52:39.90	&	5.398351917	&	$24.40 \pm 0.12$	&	$0.73 \pm 0.18$	&	...	\\
11	&	05:05:42.064	&	-67:52:39.31	&	5.49546324	&	$23.04 \pm 0.08$	&	$0.88 \pm 0.11$	&	...	\\
12	&	05:05:43.097	&	-67:52:38.26	&	5.660131952	&	$22.16 \pm 0.08$	&	$0.64 \pm 0.11$	&	...	\\
13	&	05:05:41.892	&	-67:52:46.96	&	5.758968867	&	$22.55 \pm 0.08$	&	$0.64 \pm 0.11$	&	...	\\
14	&	05:05:43.481	&	-67:52:47.47	&	5.948303175	&	$22.83 \pm 0.08$	&	$0.65 \pm 0.11$	&	...	\\
15	&	05:05:43.764	&	-67:52:43.76	&	6.004249655	&	$23.35 \pm 0.08$	&	$0.91 \pm 0.11$	&	...	\\
16	&	05:05:42.718	&	-67:52:36.80	&	6.659096267	&	$21.92 \pm 0.08$	&	$0.64 \pm 0.11$	&	...	\\
17	&	05:05:43.300	&	-67:52:37.26	&	7.061129685	&	$20.30 \pm 0.07$	&	$0.44 \pm 0.10$	&	...	\\
18	&	05:05:41.470	&	-67:52:41.74	&	7.167131162	&	$22.23 \pm 0.07$	&	$0.58 \pm 0.11$	&	...	\\
19	&	05:05:43.038	&	-67:52:50.40	&	7.190373915	&	$20.01 \pm 0.07$	&	$1.11 \pm 0.10$	&	Red Giant	\\
20	&	05:05:43.678	&	-67:52:48.10	&	7.199063542	&	$22.21 \pm 0.07$	&	$0.68 \pm 0.11$	&	...	\\
21	&	05:05:42.149	&	-67:52:36.95	&	7.22476171	&	$23.65 \pm 0.09$	&	$0.76 \pm 0.13$	&	...	\\
22	&	05:05:42.966	&	-67:52:36.33	&	7.284792615	&	$22.64 \pm 0.08$	&	$0.73 \pm 0.12$	&	...	\\
23	&	05:05:43.962	&	-67:52:40.98	&	7.533719302	&	$23.15 \pm 0.08$	&	$0.88 \pm 0.11$	&	...	\\
24	&	05:05:43.201	&	-67:52:36.45	&	7.558633919	&	$20.31 \pm 0.07$	&	$0.44 \pm 0.10$	&	...	\\
25	&	05:05:41.374	&	-67:52:42.38	&	7.577795509	&	$23.04 \pm 0.08$	&	$0.65 \pm 0.11$	&	...	\\
26	&	05:05:41.836	&	-67:52:37.58	&	7.649151977	&	$23.74 \pm 0.10$	&	$0.73 \pm 0.14$	&	...	\\
27	&	05:05:41.681	&	-67:52:38.32	&	7.724520655	&	$21.83 \pm 0.07$	&	$0.54 \pm 0.11$	&	...	\\
28	&	05:05:41.805	&	-67:52:49.31	&	7.730030699	&	$20.00 \pm 0.07$	&	$0.15 \pm 0.10$	&	...	\\
29	&	05:05:41.651	&	-67:52:48.54	&	7.805631855	&	$20.49 \pm 0.07$	&	$0.40 \pm 0.10$	&	...	\\
30	&	05:05:43.193	&	-67:52:50.96	&	7.986133622	&	$21.28 \pm 0.07$	&	$0.50 \pm 0.10$	&	...	\\
31	&	05:05:44.136	&	-67:52:43.65	&	8.100822817	&	$20.71 \pm 0.07$	&	$0.80 \pm 0.10$	&	Possible Subgiant	\\
32	&	05:05:42.540	&	-67:52:51.64	&	8.226489662	&	$21.74 \pm 0.08$	&	$1.11 \pm 0.11$	&	Subgiant	\\
33	&	05:05:44.132	&	-67:52:45.14	&	8.246537958	&	$22.51 \pm 0.08$	&	$0.72 \pm 0.11$	&	...	\\
34	&	05:05:41.213	&	-67:52:43.37	&	8.406693999	&	$23.59 \pm 0.08$	&	$0.84 \pm 0.12$	&	...	\\
35	&	05:05:42.825	&	-67:52:51.87	&	8.430785579	&	$20.60 \pm 0.07$	&	$0.50 \pm 0.11$	&	...	\\
36	&	05:05:41.410	&	-67:52:39.16	&	8.473066448	&	$21.99 \pm 0.08$	&	$0.12 \pm 0.11$	&	...	\\
37	&	05:05:41.377	&	-67:52:39.27	&	8.57772947	&	$21.79 \pm 0.08$	&	$0.00 \pm 0.11$	&	...	\\
38	&	05:05:42.102	&	-67:52:35.49	&	8.660130888	&	$23.73 \pm 0.10$	&	$0.73 \pm 0.14$	&	...	\\
39	&	05:05:42.375	&	-67:52:34.85	&	8.810468885	&	$21.84 \pm 0.07$	&	$0.40 \pm 0.11$	&	...	\\
40	&	05:05:42.062	&	-67:52:51.51	&	8.817870042	&	$22.70 \pm 0.08$	&	$0.94 \pm 0.11$	&	...	\\
41	&	05:05:41.365	&	-67:52:38.38	&	9.100873412	&	$21.83 \pm 0.08$	&	$0.52 \pm 0.11$	&	...	\\
42	&	05:05:41.161	&	-67:52:39.92	&	9.392677849	&	$21.87 \pm 0.08$	&	$0.60 \pm 0.11$	&	...	\\
43	&	05:05:44.227	&	-67:52:47.40	&	9.467368062	&	$24.69 \pm 0.14$	&	$1.13 \pm 0.19$	&	...	\\
44	&	05:05:44.286	&	-67:52:39.83	&	9.650288493	&	$21.78 \pm 0.07$	&	$0.57 \pm 0.11$	&	...	\\
45	&	05:05:43.878	&	-67:52:36.17	&	9.86504		&	$24.49 \pm 0.11$	&	$1.02 \pm 0.15$	&	...	\\
46	&	05:05:44.479	&	-67:52:42.53	&	10.07393908	&	$22.32 \pm 0.08$	&	$0.79 \pm 0.11$	&	...	\\
47	&	05:05:44.232	&	-67:52:38.11	&	10.1631294	&	$23.53 \pm 0.08$	&	$0.80 \pm 0.12$	&	...	\\
48	&	05:05:43.048	&	-67:52:53.46	&	10.18684193	&	$23.15 \pm 0.08$	&	$0.85 \pm 0.11$	&	...	\\
49	&	05:05:41.071	&	-67:52:39.08	&	10.20062612	&	$21.82 \pm 0.08$	&	$0.52 \pm 0.11$	&	...	\\
50	&	05:05:41.409	&	-67:52:36.33	&	10.20894293	&	$21.42 \pm 0.07$	&	$0.66 \pm 0.11$	&	...	\\
51	&	05:05:42.318	&	-67:52:53.63	&	10.39393569	&	$19.11 \pm 0.07$	&	$1.43 \pm 0.10$	&	Red Giant	\\
52	&	05:05:44.411	&	-67:52:39.44	&	10.45277377	&	$22.13 \pm 0.07$	&	$0.66 \pm 0.11$	&	...	\\
53	&	05:05:43.030	&	-67:52:33.17	&	10.45678128	&	$21.27 \pm 0.07$	&	$0.42 \pm 0.10$	&	...	\\
54	&	05:05:44.025	&	-67:52:50.83	&	10.48595488	&	$24.49 \pm 0.13$	&	$0.59 \pm 0.19$	&	...	\\
55	&	05:05:41.081	&	-67:52:48.83	&	10.60514694	&	$23.86 \pm 0.09$	&	$1.09 \pm 0.13$	&	...	\\
56	&	05:05:44.480	&	-67:52:39.58	&	10.76146381	&	$22.16 \pm 0.07$	&	$0.43 \pm 0.11$	&	...	\\
57	&	05:05:41.223	&	-67:52:36.42	&	10.92278639	&	$20.82 \pm 0.07$	&	$1.22 \pm 0.10$	&	Red Giant	\\
58	&	05:05:40.805	&	-67:52:45.95	&	10.99404246	&	$24.25 \pm 0.10$	&	$0.80 \pm 0.15$	&	...	\\
59	&	05:05:41.353	&	-67:52:51.99	&	11.42864292	&	$19.84 \pm 0.07$	&	$0.27 \pm 0.10$	&	...	\\
60	&	05:05:41.967	&	-67:52:32.49	&	11.72747117	&	$22.85 \pm 0.08$	&	$0.34 \pm 0.12$	&	...	\\
61	&	05:05:40.884	&	-67:52:37.60	&	11.82204936	&	$21.03 \pm 0.07$	&	$0.70 \pm 0.10$	&	...	\\
62	&	05:05:44.807	&	-67:52:42.39	&	11.93295165	&	$23.48 \pm 0.09$	&	$0.95 \pm 0.12$	&	...	\\
63	&	05:05:44.154	&	-67:52:52.23	&	11.99548438	&	$22.50 \pm 0.08$	&	$0.61 \pm 0.11$	&	...	\\
64	&	05:05:43.909	&	-67:52:33.59	&	11.99836885	&	$21.39 \pm 0.07$	&	$0.34 \pm 0.10$	&	...	\\
65	&	05:05:41.872	&	-67:52:54.62	&	12.10020096	&	$21.32 \pm 0.07$	&	$0.36 \pm 0.11$	&	...	\\
66	&	05:05:42.489	&	-67:52:31.22	&	12.2984564	&	$22.78 \pm 0.08$	&	$-0.02 \pm 0.12$	&	...	\\
67	&	05:05:41.623	&	-67:52:32.71	&	12.35798146	&	$20.81 \pm 0.07$	&	$0.33 \pm 0.10$	&	...	\\
68	&	05:05:44.837	&	-67:52:46.77	&	12.50031838	&	$22.65 \pm 0.08$	&	$0.70 \pm 0.11$	&	...	\\
69	&	05:05:40.571	&	-67:52:39.96	&	12.53115011	&	$21.73 \pm 0.07$	&	$0.44 \pm 0.11$	&	...	\\
70	&	05:05:42.334	&	-67:52:31.06	&	12.57188575	&	$22.78 \pm 0.08$	&	$0.78 \pm 0.11$	&	...	\\
71	&	05:05:41.702	&	-67:52:32.18	&	12.61488456	&	$20.67 \pm 0.07$	&	$0.32 \pm 0.11$	&	...	\\
72	&	05:05:42.767	&	-67:52:30.81	&	12.65469791	&	$23.27 \pm 0.08$	&	$0.43 \pm 0.12$	&	...	\\
73	&	05:05:44.315	&	-67:52:34.57	&	12.72994205	&	$23.43 \pm 0.09$	&	$0.62 \pm 0.13$	&	...	\\
74	&	05:05:40.693	&	-67:52:49.37	&	12.78558956	&	$22.04 \pm 0.08$	&	$0.77 \pm 0.11$	&	...	\\
75	&	05:05:43.392	&	-67:52:31.15	&	12.91198134	&	$19.63 \pm 0.07$	&	$1.22 \pm 0.10$	&	Red Giant	\\
76	&	05:05:43.953	&	-67:52:32.64	&	12.91924751	&	$21.39 \pm 0.07$	&	$0.35 \pm 0.10$	&	...	\\
77	&	05:05:43.806	&	-67:52:54.91	&	13.02700388	&	$22.79 \pm 0.08$	&	$0.57 \pm 0.11$	&	...	\\
78	&	05:05:41.477	&	-67:52:32.37	&	13.07375453	&	$20.81 \pm 0.07$	&	$-0.01 \pm 0.11$	&	...	\\
79	&	05:05:40.510	&	-67:52:39.25	&	13.07431727	&	$21.74 \pm 0.07$	&	$-0.07 \pm 0.11$	&	...	\\
80	&	05:05:40.406	&	-67:52:41.40	&	13.12516034	&	$24.26 \pm 0.12$	&	$0.67 \pm 0.19$	&	...	\\
81	&	05:05:41.846	&	-67:52:31.10	&	13.27605071	&	$19.72 \pm 0.07$	&	$0.15 \pm 0.10$	&	...	\\
82	&	05:05:41.898	&	-67:52:56.01	&	13.34051978	&	$22.95 \pm 0.08$	&	$0.60 \pm 0.12$	&	...	\\
83	&	05:05:40.580	&	-67:52:49.42	&	13.38004069	&	$21.99 \pm 0.08$	&	$0.77 \pm 0.11$	&	...	\\
84	&	05:05:40.451	&	-67:52:47.83	&	13.43881157	&	$19.80 \pm 0.07$	&	$0.06 \pm 0.10$	&	...	\\
85	&	05:05:44.286	&	-67:52:33.43	&	13.44365216	&	$21.64 \pm 0.07$	&	$0.44 \pm 0.11$	&	...	\\
86	&	05:05:44.754	&	-67:52:50.33	&	13.46306554	&	$20.60 \pm 0.07$	&	$-0.18 \pm 0.10$	&	...	\\
87	&	05:05:42.554	&	-67:52:56.99	&	13.55101806	&	$23.19 \pm 0.08$	&	$0.88 \pm 0.12$	&	...	\\
88	&	05:05:41.559	&	-67:52:31.51	&	13.58310999	&	$21.30 \pm 0.08$	&	$0.16 \pm 0.11$	&	...	\\
89	&	05:05:40.580	&	-67:52:49.88	&	13.58926613	&	$22.05 \pm 0.08$	&	$0.79 \pm 0.11$	&	...	\\
90	&	05:05:40.976	&	-67:52:52.97	&	13.61416914	&	$23.99 \pm 0.12$	&	$0.47 \pm 0.19$	&	...	\\
91	&	05:05:44.805	&	-67:52:50.16	&	13.62910757	&	$20.84 \pm 0.07$	&	$0.09 \pm 0.10$	&	...	\\
92	&	05:05:44.140	&	-67:52:54.42	&	13.63624672	&	$23.52 \pm 0.09$	&	$0.85 \pm 0.12$	&	...	\\
93	&	05:05:45.056	&	-67:52:47.53	&	13.8977275	&	$20.70 \pm 0.07$	&	$0.55 \pm 0.10$	&	...	\\
94	&	05:05:41.533	&	-67:52:55.79	&	13.97866151	&	$23.95 \pm 0.09$	&	$1.17 \pm 0.13$	&	...	\\
95	&	05:05:40.551	&	-67:52:36.52	&	13.98808514	&	$23.11 \pm 0.08$	&	$0.78 \pm 0.12$	&	...	\\
96	&	05:05:44.857	&	-67:52:50.41	&	14.00659346	&	$18.14 \pm 0.07$	&	$-0.02 \pm 0.10$	&	...	\\
97	&	05:05:45.192	&	-67:52:42.98	&	14.06992376	&	$21.01 \pm 0.07$	&	$0.68 \pm 0.10$	&	...	\\
98	&	05:05:44.339	&	-67:52:32.77	&	14.12877313	&	$21.63 \pm 0.07$	&	$0.43 \pm 0.11$	&	...	\\
99	&	05:05:40.737	&	-67:52:52.27	&	14.16630048	&	$21.63 \pm 0.07$	&	$0.40 \pm 0.11$	&	...	\\
100	&	05:05:40.862	&	-67:52:33.71	&	14.25105498	&	$23.41 \pm 0.08$	&	$1.02 \pm 0.11$	&	...	\\
101	&	05:05:40.417	&	-67:52:37.40	&	14.2591197	&	$23.27 \pm 0.09$	&	$0.57 \pm 0.13$	&	...	\\
102	&	05:05:42.989	&	-67:52:57.78	&	14.40123544	&	$21.23 \pm 0.07$	&	$0.24 \pm 0.11$	&	...	\\
103	&	05:05:40.155	&	-67:52:44.92	&	14.45493582	&	$21.27 \pm 0.08$	&	$0.26 \pm 0.11$	&	...	\\
104	&	05:05:41.245	&	-67:52:31.56	&	14.46479083	&	$23.38 \pm 0.08$	&	$0.63 \pm 0.12$	&	...	\\
105	&	05:05:40.165	&	-67:52:46.14	&	14.57446683	&	$21.28 \pm 0.08$	&	$0.50 \pm 0.11$	&	...	\\
106	&	05:05:43.808	&	-67:52:30.28	&	14.58908207	&	$22.59 \pm 0.08$	&	$0.76 \pm 0.11$	&	...	\\
107	&	05:05:41.068	&	-67:52:54.86	&	14.66494386	&	$22.87 \pm 0.08$	&	$0.73 \pm 0.11$	&	...	\\
108	&	05:05:42.655	&	-67:52:58.19	&	14.71983412	&	$23.91 \pm 0.11$	&	$0.81 \pm 0.15$	&	...	\\
109	&	05:05:45.160	&	-67:52:48.43	&	14.73838035	&	$20.70 \pm 0.07$	&	$0.55 \pm 0.10$	&	...	\\
110	&	05:05:45.264	&	-67:52:39.39	&	15.026156	&	$19.58 \pm 0.07$	&	$1.26 \pm 0.10$	&	Red Giant	\\
111	&	05:05:44.076	&	-67:52:56.48	&	15.14582152	&	$21.25 \pm 0.07$	&	$0.38 \pm 0.11$	&	...	\\
112	&	05:05:44.076	&	-67:52:56.48	&	15.14582152	&	$21.25 \pm 0.07$	&	$-0.35 \pm 0.11$	&	...	\\
113	&	05:05:40.067	&	-67:52:40.55	&	15.16244912	&	$20.48 \pm 0.07$	&	$1.16 \pm 0.10$	&	Red Giant	\\
114	&	05:05:41.520	&	-67:52:29.76	&	15.23780593	&	$21.93 \pm 0.08$	&	$0.84 \pm 0.11$	&	...	\\
115	&	05:05:43.513	&	-67:52:28.91	&	15.25098602	&	$22.58 \pm 0.08$	&	$0.76 \pm 0.11$	&	...	\\
116	&	05:05:43.158	&	-67:52:58.62	&	15.36743114	&	$21.15 \pm 0.07$	&	$0.57 \pm 0.10$	&	...	\\
117	&	05:05:45.451	&	-67:52:43.19	&	15.52580043	&	$21.49 \pm 0.07$	&	$0.31 \pm 0.11$	&	...	\\
118	&	05:05:43.281	&	-67:52:58.72	&	15.59743231	&	$21.15 \pm 0.07$	&	$0.57 \pm 0.10$	&	...	\\
119	&	05:05:39.931	&	-67:52:44.68	&	15.69375101	&	$22.04 \pm 0.08$	&	$0.64 \pm 0.11$	&	...	\\
120	&	05:05:40.292	&	-67:52:35.58	&	15.72361201	&	$23.84 \pm 0.11$	&	$0.97 \pm 0.15$	&	...	\\
121	&	05:05:40.863	&	-67:52:55.37	&	15.79999793	&	$22.99 \pm 0.08$	&	$0.75 \pm 0.11$	&	...	\\
\enddata
\label{tab:0505mags}
\end{deluxetable*}
\end{centering}

\section{LMC SNR 0509-68.7}
\label{sec:687}
\begin{figure*}
\centering
\epsscale{1.0}
\plotone{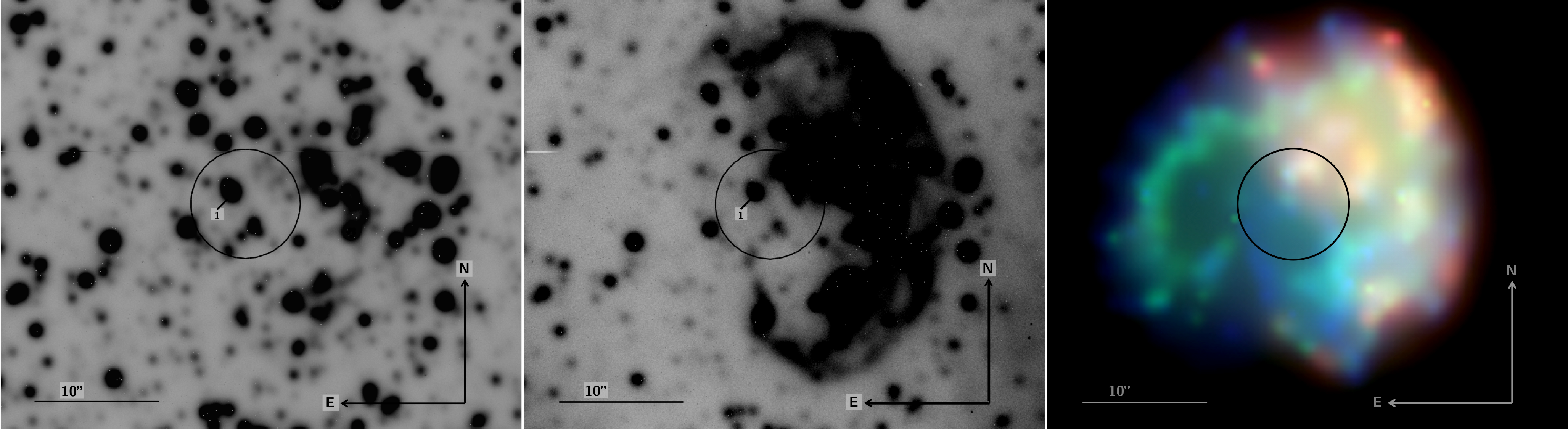}
\caption{Identically cropped images of LMC SNR 0509-687 (N103B); on the left is the combined {\it g'r'i'} Gemini image, in the center is the H$\alpha$ Gemini image, and on the right is the Chandra color image, which is a combination of 0.3-0.6 keV (red), 0.6-0.9 keV (green), and 0.9-10 keV (blue) observations. The Gemini images were taken with the GMOS imager on the 8.1m Gemini South Telescope. The 4.4$\arcsec$ 99.73\% containment central region is marked in all three images. The biggest and brightest star in the center of the remnant, marked as star 1 in the optical images, is a red giant, and therefore a possible ex-companion from a recurrent nova or symbiotic system. Additionally, there are seven other main-sequence stars inside the error circle that are bright enough to be ex-companions from supersoft X-ray sources. Because the optical (H$\alpha$) image of the remnant does not show its true extent, we used the X-ray image on the right in addition to a radio image from \citet{dickel1995a} to locate the center of the SNR. (X-ray image source: The Chandra Supernova Remnant Catalog (http://hea-www.harvard.edu/ChandraSNR/), used with permission)}
\label{fig:687triptych}
\end{figure*}

The fourth LMC Ia SNR we consider is SNR 0509-68.7 (N103B). This 860-year old remnant has X-ray spectra \citep{hughes1995a} and light echo observations (\citealp{rest2005a}; A. Rest 2010, private communication) confirming the Ia nature of the supernova associated with this remnant. Only half of the remnant is bright in H$\alpha$, which can be seen in the center panel of Figure \ref{fig:687triptych}. This asymmetry is also visible in {\it Spitzer} infrared observations, and has been identified by \mbox{{\cite{williams2014a}}} as possible circumstellar material lost from the progenitor system before the supernova explosion.

We obtained Gemini GMOS imagery of SNR 0509-68.7 in October and November of 2011 because there were no useful images in the {\it HST} archive. Because of the aforementioned asymmetry in H$\alpha$, we could not use that image to locate the geometric center of the remnant. Instead, we used radio \citep{dickel1995a} and X-ray images (right panel of Figure \ref{fig:687triptych}, \citealp{lewis2003a}; again obtained via the Chandra SNR Catalog) to locate the geometric center. Table \ref{tab:687centers} presents the measured geometric centers, as well as the explosion site and the distance any possible ex-companion stars could have traveled.

We note that the shell is almost perfectly round in radio and X-ray, with only small out-of-roundness likely caused by random variations. This implies that the explosion site should be at or very near the same location as the geometric center, so we include no formal offset between the two locations, and thus the explosion site is 05:08:59.62, -68:43:35.5 (J2000). As was the case for SNR 0519-69.0 \citep{edwards2012a}, the 1$\sigma$ uncertainty on this is approximately equal to the RMS of the shell radius measurements, which is 0.9$\arcsec$ for both radio and X-ray, so this is added into the total uncertainty on the site of the explosion. The central 99.73\% containment region for SNR 0509-68.7 has a radius of 4.4$\arcsec$; it is marked on the images in Figure \ref{fig:687triptych}.

\begin{figure}
\centering
\epsscale{1.0}
\plotone{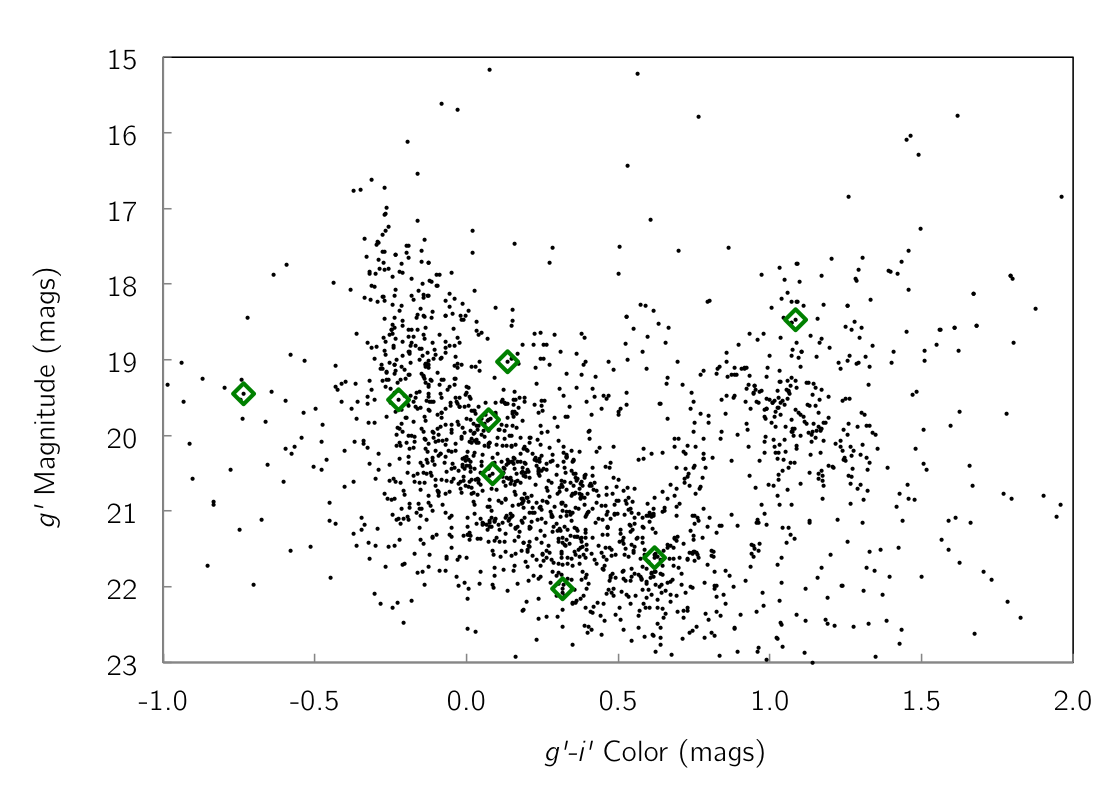}
\caption{$g'-i'$ Color-Magnitude Diagram for SNR 0509-68.7. This $g'-i'$ color-magnitude diagram constructed from our Gemini GMOS observations can be used to classify stars in the field of SNR 0509-68.7. No error bars are plotted on this figure, but a full listing of the photometry can be found in Table \ref{tab:687mags}; the average $1\sigma$ errors on $g'$ and $g'-i'$  are $\pm 0.04$ and $\pm 0.29$, respectively. The stars located within the central region of the supernova remnant are highlighted with green diamonds. There is one clear red giant, six possible main-sequence ex-companions that are bright enough to have come from supersoft X-ray sources, and one very blue star that could be the stripped core of a former red giant or a Helium star.}
\label{fig:687cmd}
\end{figure}

There are eight possible ex-companion stars located within the central containment region of SNR 0509-68.7. The bright central star is clearly a red giant based on its location on the $g'-i'$ color-magnitude diagram (Figure \ref{fig:687cmd}), there are six main-sequence stars bright enough to have come from supersoft X-ray source binaries, and there is one very blue star that could be the stripped core of a former red giant ex-companion or a Helium star. The stars located within the central error region are highlighted with green diamonds in Figure \ref{fig:687cmd} and listed in Table \ref{tab:687mags}. Any of these eight stars could be the ex-companion, which means that all SD and DD models are currently possible for SNR 0509-68.7. Again, future spectroscopic observations may be able to shed light on whether any of the stars are in fact the ex-companion star.

\begin{centering}
\begin{deluxetable*}{lllll}
\tabletypesize{\scriptsize}
\tablewidth{0pc}
\tablecaption{Positions in SNR 0509-68.7}
\tablehead{
	\colhead{Position} & 
	\colhead{RA (J2000)} & 
	\colhead{Dec. (J2000)} & 
	\colhead{Radius ($\arcsec$)} & 
	\colhead{Confidence}
	}
\startdata
Geometric center in Radio	&	05:08:59.65	&	-68:43:35.6	&	0.7	&	1$\sigma$	\\
Geometric center in X-ray	&	05:08:59.59	&	-68:43:35.3	&	0.7	&	1$\sigma$	\\
Combined geometric center of SNR	&	05:08:59.62	&	-68:43:35.5	&	0.5	&	1$\sigma$	\\
Site of explosion	&	05:08:59.62	&	-68:43:35.5	&	0.5	&	1$\sigma$	\\
Main-sequence ex-companion	&	05:08:59.62	&	-68:43:35.5	&	4.4	&	3$\sigma$	\\
\enddata
\label{tab:687centers}
\end{deluxetable*}
\end{centering}

\begin{centering}
\begin{deluxetable*}{lllllrl}
\tabletypesize{\scriptsize}
\tablewidth{0pc}
\tablecaption{Stars Inside Central Region for SNR 0509-68.7}
\tablehead{
	\colhead{Star} 				& 
	\colhead{RA (J2000)} 		& 
	\colhead{Dec. (J2000)} 		& 
	\colhead{$\Theta$ ($\arcsec$)}	& 
	\colhead{$g'$ (mag)} 			& 
	\colhead{$g'-i'$ (mag)} 	& 
	\colhead{Comments}
	}
\startdata
1	&	05:08:59.824	&	-68:43:34.54	&	1.5	&	$18.48 \pm 0.03$	&	$1.08 \pm 0.28$	&	Red Giant	\\
2	&	05:08:59.482	&	-68:43:37.30	&	1.9	&	$19.02 \pm 0.03$	&	$0.14 \pm 0.28$	&	...	\\
3	&	05:08:59.675	&	-68:43:38.12	&	2.6	&	$19.53 \pm 0.03$	&	$-0.22 \pm 0.28$	&	...	\\
4	&	05:09:00.193	&	-68:43:33.85	&	3.5	&	$20.51 \pm 0.03$	&	$0.09 \pm 0.28$	&	...	\\
5	&	05:08:58.964	&	-68:43:34.89	&	3.6	&	$19.44 \pm 0.04$	&	$-0.74 \pm 0.28$	&	Very Blue	\\
6	&	05:08:59.868	&	-68:43:39.22	&	4.0	&	$19.79 \pm 0.03$	&	$0.07 \pm 0.28$	&	...	\\
7	&	05:08:59.294	&	-68:43:31.97	&	4.0	&	$22.02 \pm 0.08$	&	$0.32 \pm 0.30$	&	...	\\
8	&	05:08:59.196	&	-68:43:38.78	&	4.0	&	$21.61 \pm 0.04$	&	$0.62 \pm 0.28$	&	...	\\
\enddata
\label{tab:687mags}
\end{deluxetable*}
\end{centering}

\section{Discussion}
\label{sec:discussion}

Many recent papers (e.g. \citealp{maoz2012a,graur2013a}) have presented compelling evidence that DDs are the currently favored progenitor channel, leading to what \mbox{\cite{maoz2014a}} describe as a ``paradigm shift" in the community, but concerns about the underlying physics remain (e.g. \mbox{{\citet{timmes1994a}}} and references therein, and discussion in \mbox{{\citet{wheeler2012b}}}). Additionally, there are strong arguments for multiple progenitor channels \citep{brandt2010a,greggio2010a,pritchet2008a}. It is therefore crucial to continue to identify the progenitors in as many systems as possible, to accumulate enough identifications that we are able to start considering the statistics of the population and find connections to other properties of the supernovae. 

One avenue to explore is whether there is a connection between the progenitor type and the star formation history of the region of the supernova. SNR 0505-67.9 is similar to our two previously published SNRs (0509-67.5 and 0519-69.0) in having very little star formation in the recent history, indicating a 72\% chance of a delayed, metal-poor progenitor. SNR 0509-68.7, however, is remarkably different from the other three, with ``vigorous" star formation in its recent past (peaking between 100 and 50 Myr ago, and again 12 Myr ago), and therefore it has a high likelihood (73\%) of being associated with a prompt/young, metal rich progenitor \citep{badenes2009a}. 

It is also possible there is a connection between progenitor classes and observed SN Ia subtypes. We have a connection between SN 1991T-like supernovae and double degenerate progenitors with LMC SNR 0509-67.5, and there appears to be a connection between Type Iax (2002cx-like) supernovae and helium novae like V445 Puppis \mbox{{\citep{mccully2014a}}}. It would not be surprising if some of the variety in observed properties of SNe is correlated with variety in the progenitor systems, but establishing any believable connection will take more than just two examples. Subtyping historical supernovae from their light echoes is challenging, but possible for at least one system (LMC SNR 0509-67.5; \mbox{{\citealt{rest2008a}}}) and hopefully for more in the future.

The progenitor system may also affect the inter- and circum-stellar medium surrounding the site of the supernova, and therefore the shape and symmetry of the expanding remnant, which provides another potentially interesting connection. For LMC SNR 0505-67.9, Spitzer 24$\mu$ imaging shows the presence of nearly symmetrically-distributed pre-existing dust, but no one has yet linked this dust directly to a progenitor candidate. For LMC SNR 0509-68.7, the pre-existing dust seen again in the Spitzer 24$\mu$ image has a dramatic effect on the optical image of the remnant (Figure {\ref{fig:687triptych}}) and has been linked to pre-supernova mass loss from a possible single-degenerate progenitor \mbox{{\citep{williams2014a}}}. At this point, the numbers are all too small to be considered for a rigorous analysis, but we encourage future studies of the possible interconnection between remnant shape, pre-existing dust, and progenitor type.

With so many possible ex-companions in each of these two remnants, it is tempting to conclude that at least one of them {\it must} be an SD ex-companion, but that is a flawed assumption. We are inherently biased toward the stars that we can see, but without further observations showing unusual features such as high radial or rotational velocities, we must not assume that we are in fact seeing the ex-companion star for either of the supernovae. We note that this problem gets worse for older remnants, as demonstrated here for the case of LMC SNR 0505-67.9, because the older the remnant is, the larger the central region in which we may find the ex-companion will be, leading to an overabundance of possible ex-companion stars. Because of this, we have many viable SD candidate ex-companions for both supernovae, especially LMC SNR 0505-67.9, and cannot rule any models out, as we have been able to do for other systems. Further observations may be able to identify SD ex-companion stars for one or both of these remnants, but we likely will not be able to definitively state that either had a DD progenitor due to the number of possible ex-companions in each remnant, an effect of their location within the bar of the LMC. At this point, the answer remains unclear, but we look forward to possible answers from future observations.

\acknowledgments
This research was supported by the National Science Foundation via grant No. AST 11-09420, the Louisiana State University Graduate School, and the Kathryn W. Davis Postdoctoral Scholar program, which is supported in part by the New York State Education Department and by the National Science Foundation under grant Nos. DRL-1119444 and DUE-1340006.

This work is based on observations obtained at the Gemini Observatory (Program ID GS-2011B-Q-30), acquired through the Gemini Science Archive, and processed using the Gemini IRAF package, which is operated by the Association of Universities for Research in Astronomy, Inc., under a cooperative agreement with the NSF on behalf of the Gemini partnership: the National Science Foundation (United States), the National Research Council (Canada), CONICYT (Chile), the Australian Research Council (Australia), Minist\'{e}rio da Ci\^{e}ncia, Tecnologia e Inova\c{c}\~{a}o (Brazil) and Ministerio de Ciencia, Tecnolog\'{i}a e Innovaci\'{o}n Productiva (Argentina). This work is also based in part on observations made by the Chandra X-ray Observatory, obtained and used with permission from the Chandra Supernova Remnant Catalog.

\end{document}